\documentclass[journal=jacsat,manuscript=communication]{achemso}
\usepackage{graphicx}
\usepackage{color}
\usepackage{titlecaps}
\usepackage{caption}
\usepackage{amssymb}
\usepackage{amsmath}
\usepackage{gensymb}
\usepackage{booktabs}
\usepackage{multicol}
\usepackage{multirow}
\usepackage[table]{xcolor}
\usepackage{etoolbox}
\usepackage{natbib}
\usepackage{tablefootnote}
\usepackage{hyperref}
\makeatletter
\def\acs@contact@details{
E-mail: \acs@email@list
}
\makeatother

\title{At-Scale Discovery of High-Voltage Cathode Active Materials for Na-Ion Batteries via Charged-Only Structures and Committee Divide-and-Conquer Learning}
\setkeys{acs}{
  abbreviations = false,
  articletitle  = true,
  biblabel      = brackets,
  biochem       = false,
  doi           = false,
  etalmode      = truncate,
  keywords      = false,
  maxauthors    = 95,
  super         = true
}
\author{Suchona Akter}
\author{Md Omar Faruque}
\author{Mohammad R. Momeni}
\email{mmomenitaheri@umkc.edu}
\affiliation{Division of Energy, Matter and Systems, School of Science and Engineering, University of Missouri--Kansas City, Kansas City 64110 Missouri, United States}
\date{}

\begin{document}

\begin{abstract}
Current cathode-active materials (CAMs) designed for sodium-ion batteries (SIBs) suffer from low operating voltages, limited energy density, irreversible phase transitions, and low intrinsic electrical conductivity. In this work, we have developed generalizable and transferable descriptor-based machine learning (ML) models, along with composition- and graph-based neural networks (NNs) and transfer learning (TL) models, all trained on a dataset of charged-only structures rather than charged/discharged pairs. Using a committee of the top four trained ML models, the mean voltage is predicted as the target property. More importantly, we show that uncertainty quantification can improve the reliability of candidate selection for training subsequent higher-generation models by including structures for which the current committee models are least confident. In this way, the new expensive \textit{ab initio} data provide information in regions where the original models most need improvement. We have showcased the transferability of the final generation models by applying them to a large database of Na-free inorganic host structures curated from Materials Project, AFLOW, OQMD, and GNoME for at-scale SIB CAM discovery. A comprehensive interactive online database, NaCAM-Atlas, was also developed, containing 14,843 CAMs with voltages greater than 3 V for SIBs. Finally, a subset of the top-ranked candidate CAMs was validated through explicit, high-throughput first-principles calculations of voltage profiles, structural robustness upon sodiation/desodiation, and electronic properties. Together, this integrated data curation, ML training and improvement, and first-principles validation strategy establishes a scalable and transferable framework for accelerating the discovery of stable, high-voltage CAMs for SIBs and beyond.
\end{abstract}


\section{1. INTRODUCTION}
\textbf{1. INTRODUCTION}\newline
Building a low-carbon society based on clean and sustainable energy has become a critical scientific challenge of the 21\textsuperscript{st} century.\cite{tarascon2001issues,armand2008building,goodenough2011challenges} In this quest for renewable energy, electrochemical energy storage plays a central role given the intermittent nature of solar and wind.\cite{he2012layered} Among available technologies to date, lithium-ion batteries (LIBs) have become the leading platform for energy storage, thanks to their high energy density, long cycle life, and high operating voltages.\cite{manthiram2020reflection,goodenough2013li,tarascon2001issues,li201830} 
Despite their commercial success, Li is considered a semi-rare element with a highly uneven distribution worldwide.\cite{you2018progress} The environmental impact of Li extraction is of significant concern as well.\cite{phogat2025comprehensive} 
Simultaneously, the rapid growth in demand for consumer electronics and stationary grid energy storage is driving the evolution of new rechargeable ion battery technologies toward lower cost, higher capacity and voltage, longer cycle life, faster charging, and improved safety, among others.\cite{song2021usability} 
In this regard, sodium-ion batteries (SIBs), which share similar electrochemistry with LIBs, offer several advantages, particularly their high abundance in nature and low cost.\cite{jeong2024machine} Na, the sixth most abundant element in the Earth's crust, is lower-cost and more widely available for large-scale electrochemical energy storage.\cite{de2016comparison}

To enable practical implementations, next-generation SIBs must combine higher energy and power densities with improved cycle life, superior safety, and reduced cost. These parameters are intrinsically linked to the chemical composition and crystal structure of the cathode active materials (CAMs). Na$^+$ ion by having a larger ionic radius (1.02 \AA) than Li$^+$ (0.76 \AA), can cause larger volume changes during charge and discharge cycles and potentially degrade structural stability.\cite{hwang2017sodium} As such, Na$^+$-hosting CAMs are often prone to phase transitions and deformation, leading to capacity decay and low power density.\cite{gupta2022understanding} Therefore, identifying and optimizing new CAMs that can reversibly intercalate/deintercalate Na$^+$ ions with high capacity and stability remains a major research focus. Extensive research efforts to date have led to the development of diverse families of CAMs for SIBs, most notably sodium layered transition-metal oxides (LTMOs, Na$_x$TMO$_2$) and polyanionic compounds.\cite{rudola2021commercialisation,bauer2018scale,broux2019high,peng2023recent,yadav2023strategies, han2015comprehensive} Unlike the octahedral (O-type) configuration in Li-ion layered LTMOs, the analogs Na-ion oxides show greater structural versatility. They are generally classified into two main groups: the O3-type, where Na ions occupy octahedral sites, and the P2-type, where Na ions reside in prismatic sites.\cite{wang2025rational,yabuuchi2012p2} P2-type oxides generally offer better structural stability, higher Na-ion conductivity, and more open channels for ion diffusion than O3 structures, resulting in lower diffusion barriers and better power density and cycling stability.\cite{lavela2024improving}  
\begin{figure*}[!t]
    \centering
    \includegraphics[width=0.99\linewidth]{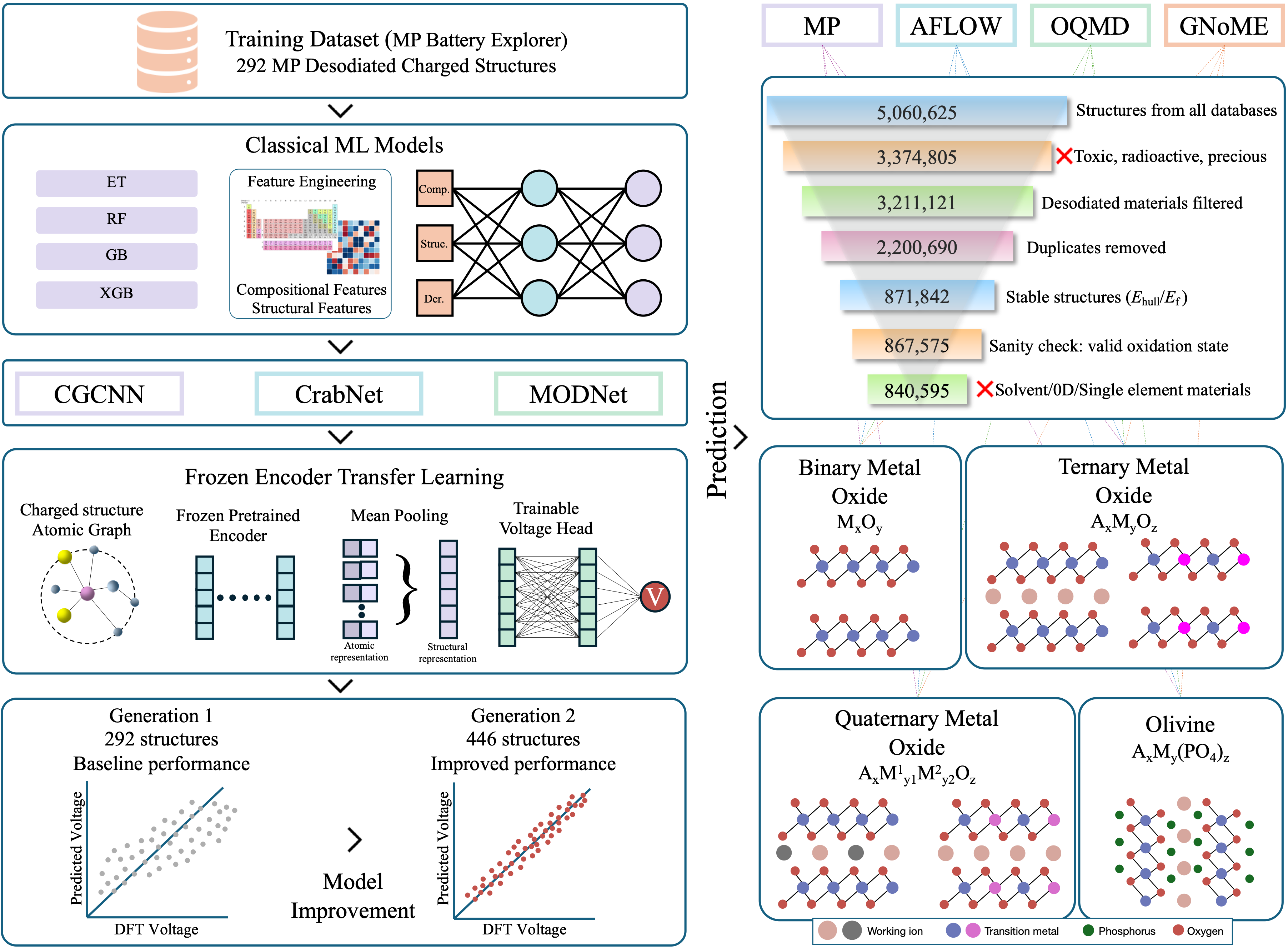}
    \caption{The overall workflow for ML developments and material screening. The left panel shows our ML model developments using structures from the Materials Project Battery Explorer dataset. They include classical models, CGCNN, CrabNet, MODNet, and frozen-encoder transfer-learning MACE and CHGNet models, as well as their improvements through subsequent 2nd generations. The right panel shows our material-screening pipeline and ML predictions for structures curated from the MP, AFLOW, OQMD, and GNoME databases.}
    \label{fig:workflow}
\end{figure*}
In addition to LTMOs, polyanionic compounds are another major family of SIB cathodes. They include olivine,\cite{ali2016polythiophene} alluaudite,\cite{trad2010namnfe2} tavorite,\cite{boivin2017vanadyl} maricite,\cite{massa2002crystal} and NASICON-type frameworks\cite{goodenough1976fast}. Polyanion chemistries include phosphates, pyrophosphates, fluorophosphates, sulfates, silicates, and mixed-polyanion compounds.\cite{li2017commercial} Strong covalent bonding within the polyanion groups generally provides good structural and thermal stability. Despite these advantages, both layered oxides and polyanionic compounds have performance limitations. For example, LTMO-based cathodes suffer from irreversible phase transitions, limited energy density, and poor air stability.\cite{acr_56_284} Polyanionic compounds, by contrast, offer greater stability but generally have lower intrinsic electrical conductivity and capacity.\cite{as_4_1600275} Therefore, further research is needed to optimize the electrochemical performance of CAMs and deliver superior properties for practical applications. 

High-throughput screening approaches using first-principles calculations have become an increasingly important tool in battery research.\cite{butler2018machine} More recently, trained machine learning (ML) models have also been integrated with high-throughput calculations to predict electrochemical properties such as voltages from structural and compositional descriptors, enabling rapid exploration of large chemical spaces that would otherwise be inaccessible through expensive density functional theory (DFT) calculations alone.\cite{sendek2017holistic} Recent works have further demonstrated the potential of pretrained machine-learning interatomic potentials (MLIPs) for voltage screening, with machine learning atomic cluster expansion (MACE) and universal models for atoms (UMA) used to estimate the energies of charged and discharged Na-ion cathode structures and thereby average intercalation voltages.\cite{thameem2026graph} However, such energy-based MLIPs and existing ML workflows still require both endpoint structures (i.e., both charged and discharged), motivating the development of models that can infer voltage directly from the charged host material alone. As mentioned, since the average intercalation voltage is obtained from the energy difference between the sodiated and desodiated host states,\cite{li2024voltage, he2024employing} pair-based models can naturally achieve high predictive accuracy. However, the requirement of both endpoint structures has significantly limited the scalability of these models. This is particularly important given that large inorganic materials databases contain mainly Na-free structures and not pairs. As a result, pair-based models cannot be readily applied across large candidate materials spaces without expensive high-throughput DFT calculations. An alternative strategy is to learn electrochemical trends directly from a single state, particularly charged or desodiated structures commonly reported in materials databases. Although removing the requirement for a paired discharged structure may reduce predictive accuracy, it broadens the applicability and scalability of material screening. By requiring only a single structural representation, charged-only models can therefore be applied to a much wider range of candidate materials, enabling rapid screening across diverse structural families where pair structures are unavailable.\cite{dinic2023unconstrained} 

In this work, we have developed and evaluated a charged-only ML framework for large-scale SIB cathode discovery. Starting from 292 desodiated electrode structures from the Materials Project (MP) Battery Explorer, we have benchmarked descriptor-based classical models, composition- and graph-based neural networks, and Frozen-encoder transfer-learning approaches based on pretrained crystal Hamiltonian graph neural network (CHGNet) and MACE models. Model generalization is evaluated using both random-split cross-validation (Random-CV) and leave-one-cluster-out CV (LOCO-CV), with the latter designed to probe extrapolation to chemically and structurally distinct materials families. We further use the prediction uncertainty from a committee of top-performing generation-1 (Gen-1) ML models to expand the training set and develop subsequent Generation-2 (Gen-2) models. We show that Gen-2-trained models achieve markedly improved accuracy when tested on an independent DFT test set of 101 structures. 
The final trained models are then applied to 840,595 Na-free structures curated from MP, AFLOW, OQMD, and GNoME databases, enabling direct charged-only voltage predictions across a broad materials space. A comprehensive interactive online platform, NaCAM-Atlas, has also been developed for exploring SIB CAMs. Finally, selected candidate materials were evaluated through explicit first-principles DFT calculations. Together, our work tests the limits and opportunities of predicting battery properties from a single structural state and offers new alternatives for large-scale materials discovery.
\\\\
 \section {2. METHODS}
 \textbf{2. METHODS}
\\\\
\subsection{2.1. Composition of the Training Dataset}
\textbf{2.1. Composition of the Training Dataset}

To construct our initial Gen-1 training dataset, we selected a subset of 416 entries from the MP Battery Explorer, with Na as the working ion. Out of this, 292 desodiated charged structures were selected. This initial training set includes a rather broad range of compositions and structures, with voltages concentrated around 2.5--4.5 V (see Supporting Information (SI) Section S1 and Figure S1 for more details). 
\\\\
\subsection{2.2. Chemical Space Partitioning Based on Composition and Structure}
\textbf{2.2. Chemical Space Partitioning Based on Composition and Structure}

\begin{figure}[t]
  \centering
  \includegraphics[width=0.99\linewidth]{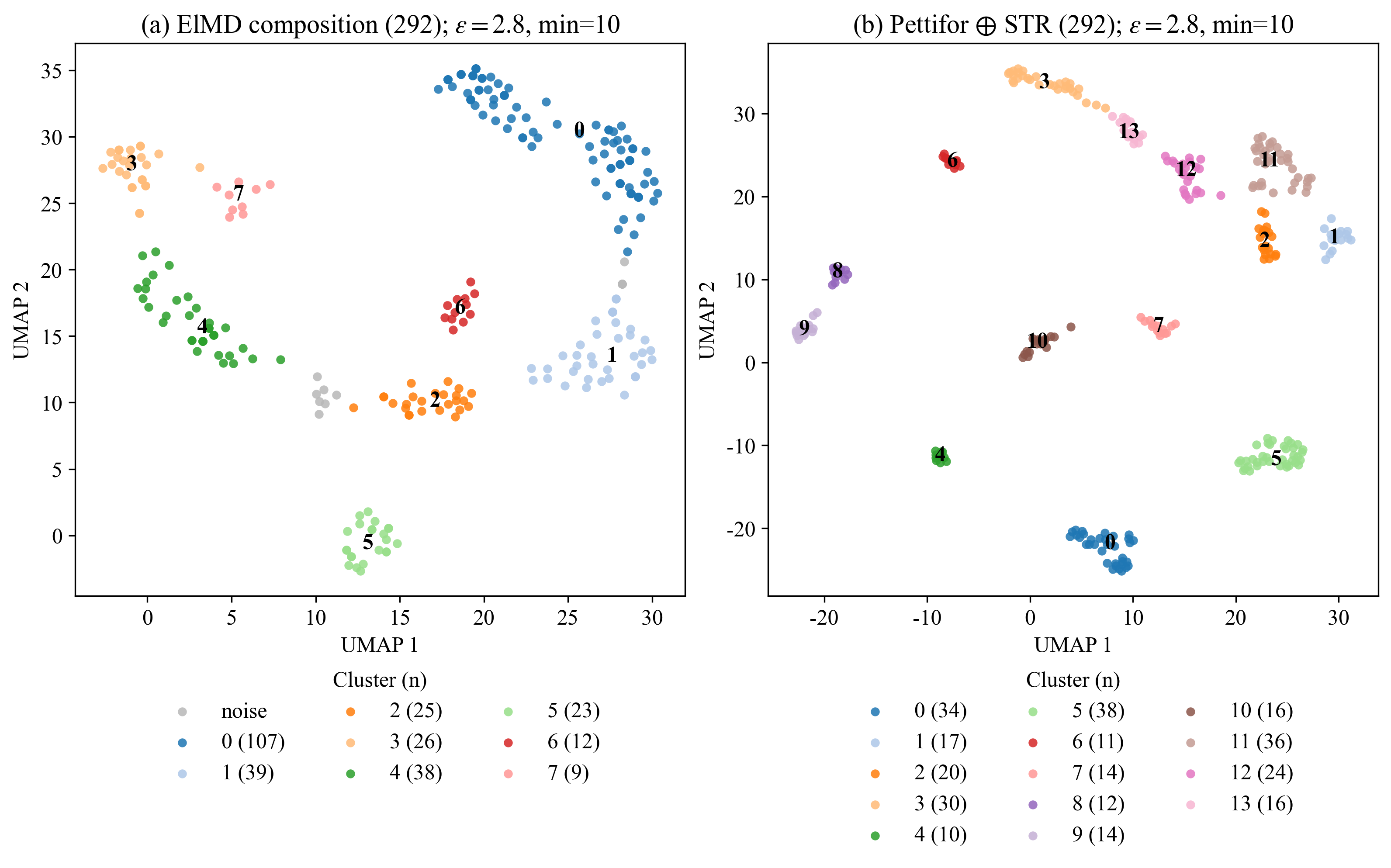}
  \caption{Visual representation of clusters generated with (a) composition-based ElMD UMAP and (b) Pettifor $\oplus$ STR UMAP. Both methods were analyzed with DBSCAN using $\epsilon=$ 2.8 and a minimum set of 10 samples.}
  \label{fig:cluster}
\end{figure}

To define chemically and structurally distinct clusters for LOCO-CV, we first partitioned the 292 charged-host structures by chemical and structural similarity. We initially used a composition-only clustering method based on Element Mover's Distance (ElMD).\cite{hargreaves2023database} In this method, each reduced chemical formula is treated as a distribution of elements. The distance between two formulas measures how much “elemental change” is needed to transform one formula into the other. This gives a useful way to compare chemical formulas without requiring crystal structures. After calculating the ElMD matrix, we used Uniform Manifold Approximation and Projection (UMAP) to map all materials onto a 2D space (Figure \ref{fig:cluster}a).\cite{healy2024uniform} 
UMAP was run with 15 neighbors, a spread of 5.0, and a random seed of 5.\cite{hargreaves2023database} Density-based spatial clustering of applications with noise (DBSCAN) algorithm\cite{ester1996density} was then used to identify dense groups, with $\epsilon=$ = 2.8 and and a minimum set of 10 samples. The ElMD map is useful because it shows broad chemical similarity between formulas. However, since ElMD only uses the reduced formula, two structures with the same reduced formula can be placed at the same point on the map, even if their crystal structures are different. This can be a problem for CAMs because voltage and Na-insertion behavior depend not only on which elements are present but also on transition-metal coordination, anion environment, local bonding, and framework geometry.
Alternatively, each charged structure was represented using a joint composition-structure descriptor that combines the Pettifor composition embedding with CrystalNN\cite{pan2021benchmarking}-based structural descriptors (Figure \ref{fig:cluster}b).
The composition representation followed the non-orthogonal Pettifor embedding introduced in Ref. \citenum{cerqueira2026non}, in which elemental vectors are arranged according to statistically derived chemical similarities rather than an orthogonal one-hot encoding. As a result, chemically similar elements are placed close together in the embedding space, so that replacing one element with a similar one causes only a small change while preserving Euclidean distance. The structural representation was built using CrystalNN local-order parameters. For each structure, 61 local-order parameters were calculated for every atom. The mean, standard deviation, minimum, and maximum of each parameter across all atoms were then computed, producing a 244-dimensional structural descriptor. This was combined with a 98-dimensional Pettifor composition descriptor, resulting in a final 342-dimensional feature vector. Before combining them, the composition and structural descriptors were each L2-normalized and multiplied by $\sqrt{0.5}$. This ensured that both descriptors contributed equally when calculating the squared Euclidean distance used for dimensionality reduction and clustering. The combined descriptors were then reduced to 2D using UMAP with 15 neighbors, a spread of 5.0, and a random seed of 5. Finally, DBSCAN was applied for clustering using an $\epsilon$ value of 2.8 and a minimum of 10 samples per cluster. The cluster sizes were 34, 17, 20, 30, 10, 38, 11, 14, 12, 14, 16, 36, 24, and 16. This combined representation, shown in Figure \ref{fig:cluster}b, was selected for LOCO-CV as it includes both compositional and structural similarity.
\\\\
\subsection{2.3. Feature Engineering}
\textbf{2.3. Feature Engineering}

Compositional and structural descriptors were generated for the 292 Na-free charged endpoint structures obtained from the MP battery dataset. For each charged/discharged pair, only the desodiated charged crystal structure was used to generate the features. The corresponding average voltage from the battery pair metadata was used as the supervised learning target. Crystal structures were processed using \textsc{pymatgen}, with additional descriptors generated using \textsc{matminer}.\cite{ward2018matminer}
Composition-based descriptors describe the elements present in each structure. These included the fractions of TMs and anions, as well as composition-weighted statistics of elemental properties such as atomic mass, Pauling electronegativity, atomic radius, covalent radius, Mendeleev number, periodic-table row, and periodic-table group. We also included descriptors that describe the chemical contrast between TM and anions, such as TM and anion electronegativities, electronegativity differences, ionic-size differences, and TM-to-anion radius ratios.\cite{ward2016general}
The structure-based descriptors describe the geometry of each material. These included density, unit-cell volume, volume per atomic site, lattice constants, lattice angles, lattice aspect ratios, and lattice anisotropy. Crystal symmetry was included using one-hot encoded crystal-system labels. Local TM environments were described using coordination-number statistics and TM-anion bond-length statistics, calculated with a 3.5~\AA~neighbor cutoff.
Additional descriptors from \textsc{matminer} were also included to provide compact descriptions of chemical bonding and structural organization. These included valence-orbital descriptors, which describe the average and fractional contributions of \emph{s}-, \emph{p}-, and \emph{d}-valence electrons. We also included structural descriptors related to structural complexity, chemical ordering, packing efficiency, bond-length distortion, variation in neighbor distances, and local structural heterogeneity. Together, these descriptors provide a compact representation of each material's chemistry and structure. All features were standardized via median imputation and standard scaling.
\\\\
\subsection{2.4. Classical Regression Models}
\textbf{2.4. Classical Regression Models}

Classical regression models were implemented using scikit-learn,\cite{pedregosa2011scikit} with XGBoost implemented using the XGBoost package.\cite{chen2016xgboost} Tree-based ensemble models, including gradient boosting (GB), random forest (RF), extremely randomized trees (ET), histogram-based gradient boosting (HGBDT), and XGBoost (XGB), were considered alongside regularized linear models and kernel-based regressors.\cite{qian2025feature} These models work especially well when limited labeled data are available.\cite{he2024employing} Earlier battery-screening studies have shown that ensemble and descriptor-based models often outperform graph-based deep learning models when labeled training data are limited.\cite{he2024employing} As we will show later, this is the case in this work as well, as our initial training dataset is limited to 292 charged structures.  

Model performance was evaluated using both Random-CV and LOCO-CV. CV evaluates a model's ability to predict unseen samples not used during model fitting. In random \(k\)-fold CV, the dataset is randomly partitioned into \(k\) subsets. In each iteration, \(k-1\) folds are used for training while the remaining one fold is used for validation, and the procedure is repeated until each fold has served at least once as the validation set.\cite{wu2026machine}
Because random splitting can place chemically and structurally similar materials in both the training and validation folds, it primarily evaluates interpolation within the sampled materials space. On the other hand, LOCO-CV more stringently assesses extrapolation to chemically and structurally distinct materials families. For LOCO-CV, 14 composition-structure clusters defined in Section 2.2 were used. In each LOCO iteration, one complete cluster was withheld for validation while the model was trained on the remaining 13 clusters. This procedure was repeated until each cluster had served once as the hold-out validation set. For direct comparison, a 14-fold Random-CV was also performed using the same dataset. The four best-performing classical models, which include ET, RF, GB, and XGB, were then selected based on their CV MAEs to form a final classical ensemble model.

\subsection{2.5. CGCNN, CrabNet, and MODNet Models}
\textbf{2.5. CGCNN, CrabNet, and MODNet Models}
The crystal graph convolutional neural network (CGCNN)\cite{xie2018crystal}, compositionally-restricted attention-based network (CrabNet)\cite{wang2021compositionally,wang2022crabnet}, and material optimal descriptor network (MODNet)\cite{de2021materials} models were also trained using the same 292 structure dataset. For CGCNN, the structures were converted to crystal graphs using a 6.0 \AA~ neighbor cutoff and a maximum of 12 neighbors per atom. 64-dimensional atom features were used with three graph-convolution layers, a 128-unit hidden crystal representation, and two hidden layers after pooling. Adam optimization was used with a learning rate of $10^{-3}$, L1 loss, batch size of 32, and a maximum of 300 epochs, with early stopping after 30 epochs without improvement in validation MAE. 
CrabNet was trained from scratch as a composition-only model using reduced chemical formulas. Elemental inputs were represented using the mat2vec\cite{tshitoyan2019unsupervised} embeddings and processed through the model's attention-based composition representation, with a base learning rate of $10^{-4}$, maximum learning rate of $6\times10^{-3}$, batch size of 128, and up to 500 epochs. 
Finally, we trained MODNet using 150 compositional descriptors together with 260 structural descriptors. The compositional descriptor pool included stoichiometric, Magpie elemental property, valence orbital, ionic property, and transition metal features. In contrast, the structural descriptors included density, symmetry, packing, structural heterogeneity, chemical ordering, and CrystalNN descriptors.

\subsection{2.6. Transfer learning with pre-trained graph neural networks}
\textbf{2.6. Transfer learning with pre-trained graph neural networks}

Transfer learning (TL) was performed using CHGNet and MACE-MPA-0 medium models as pretrained crystal-structure encoders. CHGNet is a graph neural network pretrained on energies, forces, stresses, and magnetic moments from $\sim$ 1.58 million structures in the MP trajectory dataset.\cite{deng2023chgnet} MACE models are equivariant graph neural networks that learn local atomic environments through message passing and higher-order interactions.\cite{batatia2022mace} MACE-MPA-0 and MACE-MP-0b3 used in this work belong to the larger family of pretrained universal machine-learning interatomic potentials (MLIPs) based on MACE, designed to predict atomic energies and forces for atomistic simulations and materials chemistry.\cite{batatia2025Foundation} The same training sets were used for our TL. We did not use the pretrained models' original energy outputs to predict voltage. Instead, atomic features from the pretrained network were extracted and converted into a single representation for each crystal. The atomic features were then averaged over all atoms using mean pooling. The pooled representation was passed to a multilayer perceptron head for voltage prediction. The prediction head contained a hidden layer with 64 units, a SiLU activation function, dropout with a probability of 0.1, and a single scalar output representing the average voltage.

The final models were trained using five different seed values. Predictions from the five models were averaged to obtain the final ensemble prediction for each structure.
During head-only transfer learning, all pretrained CHGNet parameters were frozen, and only the newly added voltage-prediction head was optimized. The head was trained using AdamW and smooth L1 loss on standardized voltage targets, with a learning rate of $1\times10^{-3}$, weight decay of $1\times10^{-4}$, and batch size of 8. Training was limited to 100 epochs. The checkpoint with the lowest validation MAE in volts was retained, with early stopping after 15 consecutive epochs without an improvement greater than $1\times10^{-6}$ V.
The pretrained MACE model was used as a frozen structural encoder. Invariant scalar atomic features from its final product layer were mean-pooled over all atoms to obtain a fixed-length representation for each crystal. Only the newly added voltage-prediction head was optimized. The head was trained using AdamW and smooth L1 loss on standardized voltage targets, with a learning rate of $1\times10^{-3}$, weight decay of $1\times10^{-4}$, and a batch size of 4. Training was limited to 100 epochs. The checkpoint with the lowest validation MAE in volts was retained, with early stopping after 15 consecutive epochs without an improvement greater than $1\times10^{-6}$ V.
\\\\
\subsection{2.7. Uncertainty-Guided Development of the Generation-2 Models}
\textbf{2.7. Uncertainty-Guided Development of the Generation-2 Models}

All Gen-1 voltage models were initially trained using 292 charged structures from the MP battery dataset. These models were then used to predict voltages for all the curated structures (see section 2.10 for details). To decide which structures to evaluate next with DFT, all candidates were ranked by the standard deviation of the predicted voltages across the classical models. From the most uncertain regions, 154 MP structures were selected, and their voltages were calculated using high-throughput DFT. These 154 newly labeled structures were then added to the original 292 Gen-1 structures, increasing the dataset to 446 structures. This expanded dataset was subsequently used to train the Gen-2 voltage models. The Gen-2 dataset was repartitioned using the Pettifor $\oplus$ STR representation, with $\epsilon=3.3$ and a minimum cluster size of 10. The resulting clusters are shown in SI Figure S2.

\subsection{2.8. Independent Test Set Construction}
\textbf{2.8. Independent Test Set Construction}
To evaluate the accuracy of Gen-1 and Gen-2 models, an independent test set was constructed from the combined pool of curated structures from MP, AFLOW, OQMD, and GNoME. 
\begin{figure}[h]
  \centering
  \includegraphics[width=0.99\linewidth]{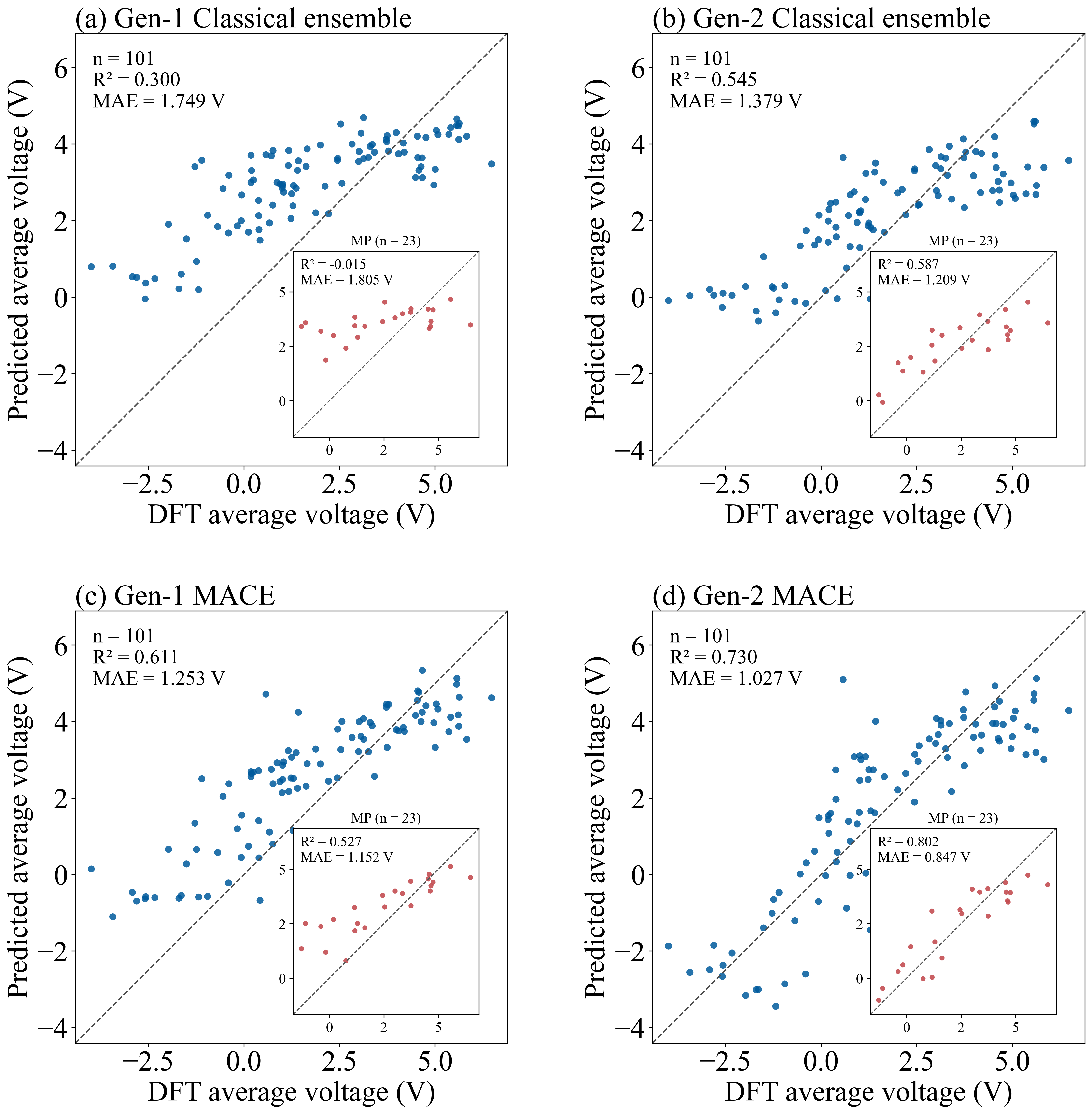}
  \caption{Correlation of DFT-calculated average voltages with average values predicted from Gen-1 and Gen-2 classical ensemble (a and b) vs. MACE (c and d) models. Results are shown for the complete 101-structure independent test set and for the MP subset with 23 structures in the insets.}
  \label{fig:correlation}
\end{figure}
To build this test set, the Gen-1 classical-ensemble model was first used to predict the voltages. The mean predicted voltages and intermodel standard deviations across the four classical models were then used to identify the final test candidates. To ensure that the test set samples a broad voltage range rather than only high-voltage candidates, 25 structures were collected across six voltage ranges for DFT calculations, yielding 150 initially selected candidates. DFT calculations converged successfully for 101 structures, which constituted our final independent test set. None of these 101 structures was included in either the Gen-1 or Gen-2 training data. The final test set contains structures from all four source databases: MP = 23, OQMD = 41, GNoME = 22, and AFLOW = 15.
\\\\
\subsection{2.9. DFT Validations}
\textbf{2.9. DFT Validations}

To assess the reliability of our ML models' predictions and also expand our training dataset, high-throughput DFT calculations were performed using \textsc{atomate2}\cite{ganose2025atomate2} together with the Vienna \textit{ab initio} simulation package (\textsc{VASP} 6.4.3)\cite{kresse1993,kresse1994,kresse1996,kresse1996efficient}. Standardized parameters were adopted from MP, including the Perdew-Burke-Ernzerhof (PBE) exchange-correlation functional\cite{perdew1996generalized}, $k$-point density, Hubbard U correction, and 520 eV CUTOFF. The Na-insertion workflow is given in the SI Figure S3. Initially, the host structures were fully relaxed using the MP relaxation workflow as implemented in \textsc{atomate2}. Next, favorable and unique Na insertion sites were identified using the charge-density insertion algorithm\cite{shen2020charge}. In this scheme, Na ions are inserted into low-electron-density regions identified by DFT calculations. Both atomic positions and cell vectors of the resulting structures were subsequently fully relaxed to determine energetically favorable sodiated configurations. Two filtering criteria were applied during our Na-insertion calculations. First, the crystal lattice of the inserted structure had to remain structurally similar or topotactic to the original desodiated host structure. This was checked using \textsc{atomate2}, which utilizes \textsc{pymatgen}'s StructureMatcher.\cite{kim2024first} Candidates were discarded if no topotactic Na insertion structure could be obtained. Second, structures were removed if the volume increased by more than 20\% compared to the relaxed Na-free host material.\cite{pascazio2025toward} This filtering criterion is applied because large volume changes are known to lead to structural deformations and low cycling performance in SIB cathodes. As mentioned previously, this is especially important for SIBs because Na$^{+}$ ions are larger than Li$^{+}$ and can therefore cause more lattice distortions during ion intercalation/deintercalation.
The Na insertion voltages were calculated from the total energies of the fully relaxed charged and discharged structures using the equation:
\begin{equation}
V = - \frac{E(\mathrm{Na}_{x_2}\mathrm{Host}) - E(\mathrm{Na}_{x_1}\mathrm{Host}) - (x_2 - x_1)E(\mathrm{Na})}{x_2 - x_1}
\end{equation}
where $E(\mathrm{Na}$) is the DFT energy per atom of the metallic sodium, and $x_1$ and $x_2$ represent the Na stoichiometries before and after insertion, respectively. Band gaps were calculated from spin-polarized Kohn–Sham eigenvalues as the energy difference between the conduction-band minimum (CBM) and valence-band maximum (VBM) over the $k$-point mesh.
\\\\
 \subsection{2.10. Database Construction and Curation}
\textbf{2.10. Database Construction and Curation}\newline
Our data curation workflow is shown in Figure \ref{fig:workflow}. For data-driven research to yield meaningful results, reliable and high-quality data are crucial. To build a dedicated materials database useful for battery cathode discovery, we began with the MP database.\cite{ong2015materials} The structures in this database come from the inorganic crystal structure database (ICSD) or data-mined chemical substitution algorithms.\cite{liu2019high} We accessed this database through the Materials Application Programming Interface (MAPI) \cite{ong2015materials} via \textsc{pymatgen}. We also retrieved crystal structures from OQMD using its OPTIMADE API\cite{evans2024developments} endpoint through the pymatgen.ext.optimade.OptimadeRester client. Additionally, structures were mined from the AFLOW repository using the AFLUX REST API\cite {rose2017aflux} and from the GNoME database via its Google Cloud Storage\cite{merchant2023scaling}.


All collected materials from the above-mentioned databases were processed using \textsc{pymatgen} to parse them into a consistent object representation and to compute their key descriptors. The structures were curated with the following steps: (a) toxic, radioactive, and precious-metal-containing materials were removed. These included As, Au, Ir, Pt, Pd, Rh, Tc, and atoms with atomic number greater than 83.\cite{pascazio2025toward} (b) Any structure with Na was removed from the pool. (c) \textsc{pymatgen}'s StructureMatcher function was then used to remove duplicate structures, ensuring that unique systems were included in subsequent high-throughput calculations.\cite{zhou20192dmatpedia} The default tolerance parameters were used for fractional length (ltol= 0.2), site position (stol= 0.3), and relative angles (angletol= 5$^\circ$).\cite{won2025mapping} (d) Unstable materials with $E_\mathrm{hull}\ge 0.1$ eV/atom when available, otherwise with positive formation energy, were identified and removed.\cite{li2024voltage, chen2025ai} (e) The oxidation states of all unique structures were then calculated using \textsc{pymatgen}'s BVAnalyzer module with default settings\cite{ong2013python,o1991atom,dembitskiy2025benchmarking}, based on local bonding environments. When the bond valence analysis failed or yielded ambiguous results, an oxidation-state analyzer based on compositional rules and the statistical distribution of oxidation states in ICSD was used.\cite{li2024voltage} (f) Next, all materials were screened to remove non-extended (i.e., 0D) structures based on structural dimensionality. For this purpose, a crystal graph was generated using the CrystalNN near-neighbor algorithm. The dimensionality of the structure was then evaluated using the Larsen method.\cite{larsen2019definition} In this analysis, each component is assigned an integer dimensionality of 0, 1, 2, or 3, where a 0D component corresponds to an isolated cluster. Such 0D materials were removed from our database. Finally, any structure containing solvent molecules was removed from our final curated database.
\\\\
\section{3. RESULTS AND DISCUSSION}
\textbf{3. RESULTS AND DISCUSSION}\newline

\subsection{3.1. Predictive Performance of Gen-1 Models}
\textbf{3.1. Predictive Performance of Gen-1 Models}

The performance of all considered Gen-1 models under LOCO-CV and Random-CV is summarized in Table~\ref{tab:model_comparison}. As mentioned previously, Random-CV measures interpolation within the sampled chemical space, whereas LOCO-CV more stringently tests chemical extrapolation by withholding entire unique chemical clusters during training. Under LOCO-CV, CHGNet and MACE-MPA-0 (MACE from now on) models achieved the lowest MAEs (0.70 and 0.71 V, respectively), compared with 0.79--0.84 V for the four classical models. In contrast, under Random-CV, the classical models achieved lower MAEs of 0.54--0.57 V, compared with 0.61 V for CHGNet and 0.59 V for MACE. This indicates that classical descriptor-based models perform particularly well for interpolation, whereas the TL CHGNet and MACE models generalize better to chemically distinct regions of the dataset. Based on their complementary interpolation and extrapolation performances, we created a committee of the four classical models and carried it forward, along with CHGNet and MACE, for the development of the Gen-2 models.
\begin{table}[htbp]
\centering
\resizebox{\columnwidth}{!}{%
\begin{tabular}{lcc}
  \toprule
  \textbf{Model} & \textbf{LOCO-CV MAE} & \textbf{Random-CV MAE} \\
                 & \textbf{(V)} & \textbf{(V)} \\
  \midrule
  Extra Trees              & 0.83 & 0.54 \\
  Random Forest            & 0.82 & 0.57 \\
  Gradient Boosting        & 0.84 & 0.55 \\
  XGBoost                  & 0.79 & 0.54 \\
  Mean Classical Models    & 0.82 & 0.55 \\
  CGCNN                     & 1.01 & 0.72 \\
  CrabNet                   & 0.99 & 0.63 \\
  MODNet                   & 1.03 & 0.60 \\
  \textbf{MACE-MPA-0 (TL)}  & \textbf{0.71} & \textbf{0.59} \\
  MACE-MP-0b3 (TL)         & 0.85 & 0.68 \\
  CHGNet (TL)              & 0.70 & 0.61 \\
  \bottomrule
\end{tabular}%
}
\caption{Comparison of Gen-1 model performance under LOCO-CV with 14 clusters and 14-fold random CV. Lower MAE indicates better performance.}
\label{tab:model_comparison}
\end{table}
\\\\
\subsection{3.2. Uncertainty-Guided Training-Set Expansion and Independent Validation}
\textbf{3.2. Uncertainty-Guided Training-Set Expansion and Independent Validation}

As described in Section~2.7, uncertainty among the top ML model predictions was used as a selection criterion to identify additional structures from regions where the Gen-1 models were least confident. For all Gen-2 trainings, DFT-calculated voltages for 154 selected structures were added to the original training data, expanding the voltage dataset from 292 to 446 structures. Figure~\ref{fig:correlation} compares classical and MACE predictions from the Gen-1 and Gen-2 models against DFT voltages for this independent test set. The classical ensemble prediction improves from Gen-1 to Gen-2, with the MAE decreasing from 1.749 to 1.379 V for the full 101-structure test set and from 1.605 to 1.209 V for the 23-structure MP subset. MACE achieves a higher predictive accuracy. For the full 101-structure test set, the MACE model MAE decreases from 1.253 V for Gen-1 to 1.027 V for Gen-2, an absolute reduction of 0.226 V and a relative improvement of about 18\%. For the 23-structure MP subset, the MAE decreases from 1.152 to 0.847 V, corresponding to an absolute reduction of 0.305 V, or approximately 26\%. The increase in $R^2$ further indicates that Gen-2 models better reproduce the variation in DFT voltages across the independent test set. The larger improvement for the MP subset is consistent with the Gen-2 training set, which includes additional structures from MP. CHGNet shows a similar improvement following Gen-2 training-set expansion, with reduced prediction errors for both the entire independent test set and the MP subset (see SI Figure S4). As shown in SI Table S1, the Random-CV and LOCO-CV MAEs are higher for the Gen-2 models than for Gen-1. However, these values do not represent a direct comparison of model performance. The Gen-2 dataset contains 154 additional structures deliberately selected from regions of high Gen-1 uncertainty, making the dataset more challenging to predict. This behavior is especially apparent under LOCO-CV, where an entire chemical cluster is withheld from training. The fixed independent test set provides a robust assessment of the effect of training-set expansion because both generations evaluate the same structures. The reduction in independent-test MAE and increase in $R^2$, despite higher internal validation errors, indicate that uncertainty-guided expansion increases coverage of difficult regions of chemical space and improves prediction on previously unseen structures.
\\\\
\subsection{3.3. Large-Scale Screening of the Curated Materials Space}
\begin{figure}[!h]
  \centering
  \includegraphics[width=0.99\linewidth]{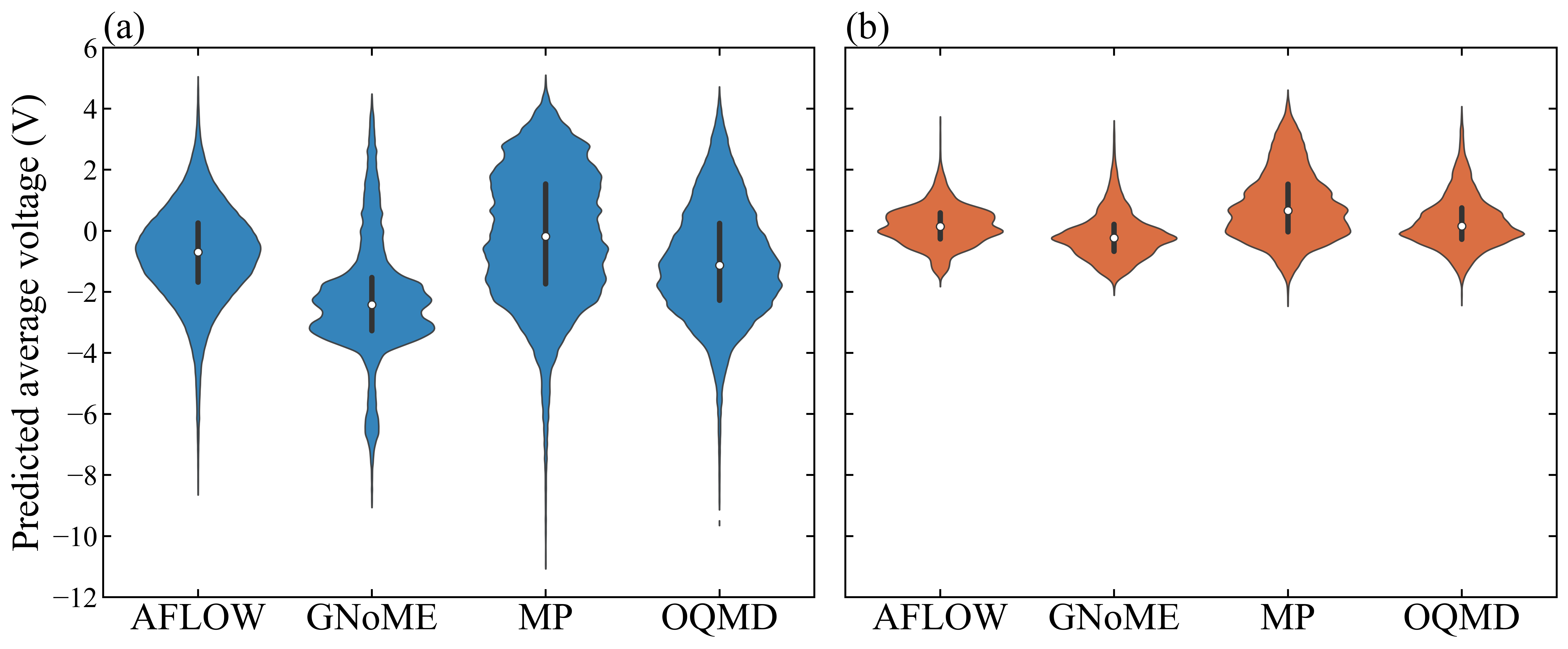}
  \caption{Predicted average-voltage distributions for the curated structures from AFLOW, GNoME, MP, and OQMD using (a) the Gen-2 MACE model and (b) the mean prediction from the classical-ensemble model. The different distribution widths illustrate the model dependence of extrapolative voltage predictions across the large curated materials space.}
  \label{fig:voltage_dist}
\end{figure}

\textbf{3.3. Large-Scale Screening of the Curated Materials Space}
Our final developed MACE and classical-ensemble Gen-2 voltage models were next applied to our full curated materials space, including 840,595 Na-free host structures collected from MP, AFLOW, OQMD, and GNoME after applying the chemical, structural, and stability filters described in Section~2.10 (see Figure~\ref{fig:workflow}). This curated database contains many crystal structures with compositions that are poorly represented or absent in the training set. These predictions, therefore, involve substantial chemical extrapolation, particularly for structures originating from AFLOW, OQMD, and GNoME. Each structure was represented using the same input representation used during model development. The classical models received the composition and structural descriptors, whereas MACE received the structure through structural representation. The predictions from the ET, RF, GB, and XGB models were combined into a four-model classical ensemble, and their mean predicted voltage was used for ranking. We note that averaging across models reduces unusually high or low predictions from any individual model and stabilizes screening of the large candidate pool. Figure~\ref{fig:voltage_dist} shows a rather strong database-dependent difference in the predicted voltage distributions. 
It also highlights a substantial difference between MACE and the classical ensemble predictions. The classical ensemble yields a relatively compact distribution centered near zero, whereas MACE predicts a substantially broader distribution extending to both positive and strongly negative voltages. For both model types, the MP population shifts toward higher predicted voltages relative to the other databases and exhibits strong positive-voltage tails. AFLOW is shifted toward lower voltages than MP in both models. GNoME shows the strongest shift toward low predicted voltages, particularly for MACE, where the central portion of the distribution lies well below zero, and a pronounced negative-voltage tail is observed. The classical ensemble predicts a much narrower GNoME distribution, although its center also remains below that of MP. OQMD occupies an intermediate regime. Its classical-ensemble distribution is centered near zero with a positive-voltage tail, whereas MACE predicts a broader distribution shifted toward negative voltages (Figure~\ref{fig:voltage_dist}).

As mentioned above, a notable feature of Figure~\ref{fig:voltage_dist} is the substantially broader prediction range produced by MACE. Across all four databases, MACE generates wider voltage distributions and more extreme negative predictions than the classical ensemble. In contrast, averaging the four classical models produces considerably more compact distributions. This divergence is particularly pronounced for GNoME and OQMD. It indicates that the two model classes respond differently when extrapolated beyond the chemical environments represented in the relatively small training set. The negative-voltage tail predicted by MACE is dominated by borides, borocarbides, and phosphides that are poorly represented in the training set. For example, several BN, VB$_2$, metal borocarbide, and higher-boride structures receive MACE predictions below $-10$ V, despite having chemically reasonable crystal structures. This trend is qualitatively consistent with previous SIB studies. Several TM borides and borocarbides have been investigated for anode materials rather than potential cathodes.\cite{ma2020boride,abdullahi2021borocarbide} However, the magnitudes of the negative tail can be attributed to out-of-domain extrapolation. We note that MACE foundation models and their fine-tuned versions have shown large extrapolation capabilities when evaluated outside the training domain.\cite{radova2025finetuning,deng2025Systematic} In contrast, tree-based classical models produce narrow distributions because they are known for poor extrapolation outside their training domain.\cite{hatakeyama2021, li2026} Therefore, their more compact distributions do not automatically make them more accurate for these extrapolative materials. Rather, the contrasting distributions illustrate different extrapolation characteristics of the two model classes.
\\\\
\subsection{3.4. Chemical-Family Trends in Predicted Voltage}
\begin{figure}[t]
  \centering
  \includegraphics[width=0.99\linewidth]{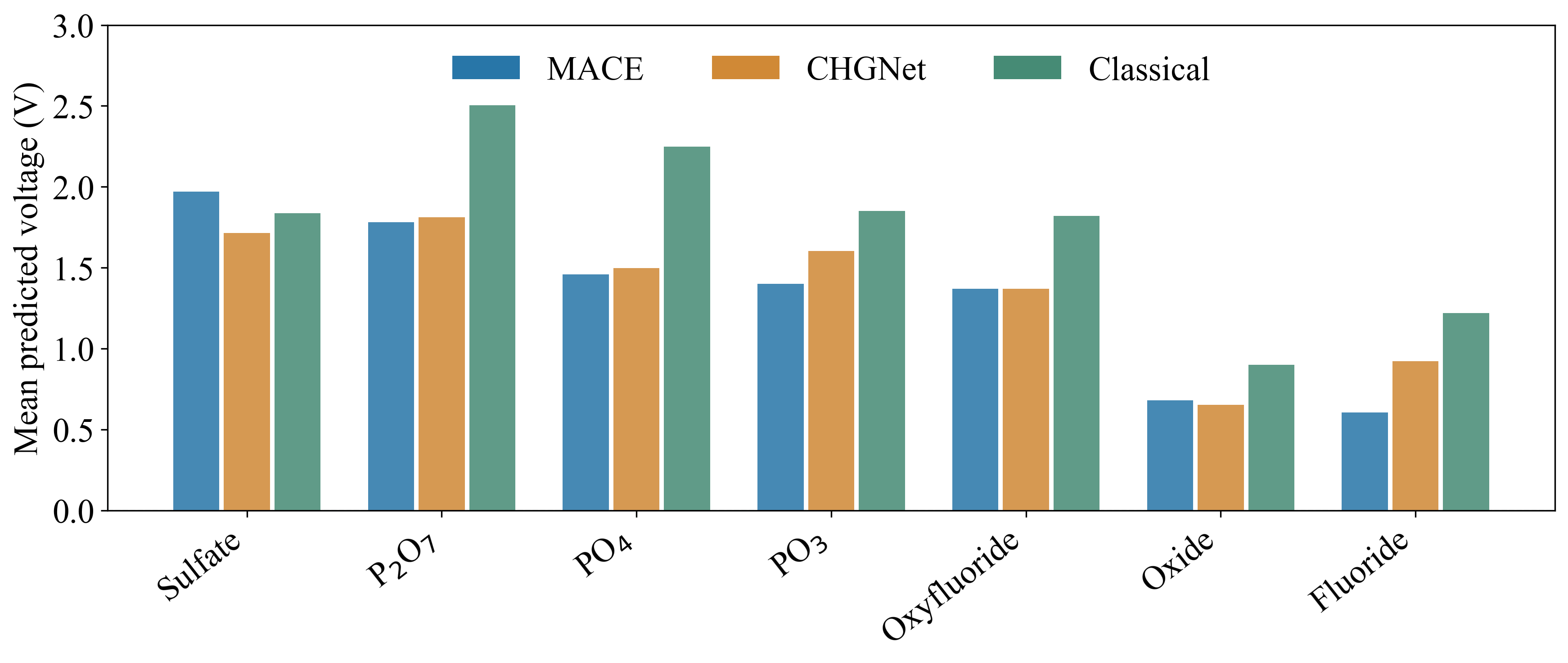}
  \caption{Mean predicted voltages for major anion and polyanion families from MACE, CHGNet, and the classical-ensemble models.}
  \label{fig:chemical_family}
\end{figure}

\textbf{3.4. Chemical-Family Trends in Predicted Voltage}
The mean predicted voltages for the dominant anion and polyanion families were analyzed to investigate possible trends (Figure \ref{fig:chemical_family}). 
Despite their different representations and extrapolation characteristics, all three MACE, CHGNet, and classical-ensemble models recover a pronounced dependence of predicted mean voltages on anion chemistry. Polyanionic families, including sulfate, pyrophosphate (P$_2$O$_7$), phosphate (PO$_4$), and metaphosphate-like PO$_3$ environments, generally show higher predicted voltages than the oxide family. This overall trend is consistent with the well-established inductive effect in polyanionic cathode materials.\cite{ling2021phosphate,jin2020polyanion} Strong covalent X–O bonding in polyanion groups influences the TM-oxygen bonding and shifts the TM redox couple to lower energy relative to Na/Na$^{+}$, resulting in a higher operating voltage. Among the different families shown in Figure \ref{fig:chemical_family}, the high sulfate voltage predicted by MACE is particularly consistent with Na-ion electrochemistry. Sulfate groups exert a stronger inductive effect than many phosphate-based polyanions.\cite{barpanda20143} The oxyfluoride family provides another useful chemical comparison. All three models predict higher mean voltages for oxyfluorides than for simple oxides. This trend can be attributed to the high electronegativity of F, which increases the ionic character of the local TM environment and shifts TM redox couples toward higher potentials.\cite{deb2024exploration} However, a different trend is observed for the pure-fluoride family, where the three models show greater variation, with, for example, MACE predicting slightly lower voltages than the oxide family. This does not necessarily contradict the expected electronegativity trend, as the family averages include different TMs, oxidation states, and structural environments rather than a direct O-to-F substitution. 
\begin{figure*}[t]
  \centering
  \includegraphics[width=0.9\linewidth]{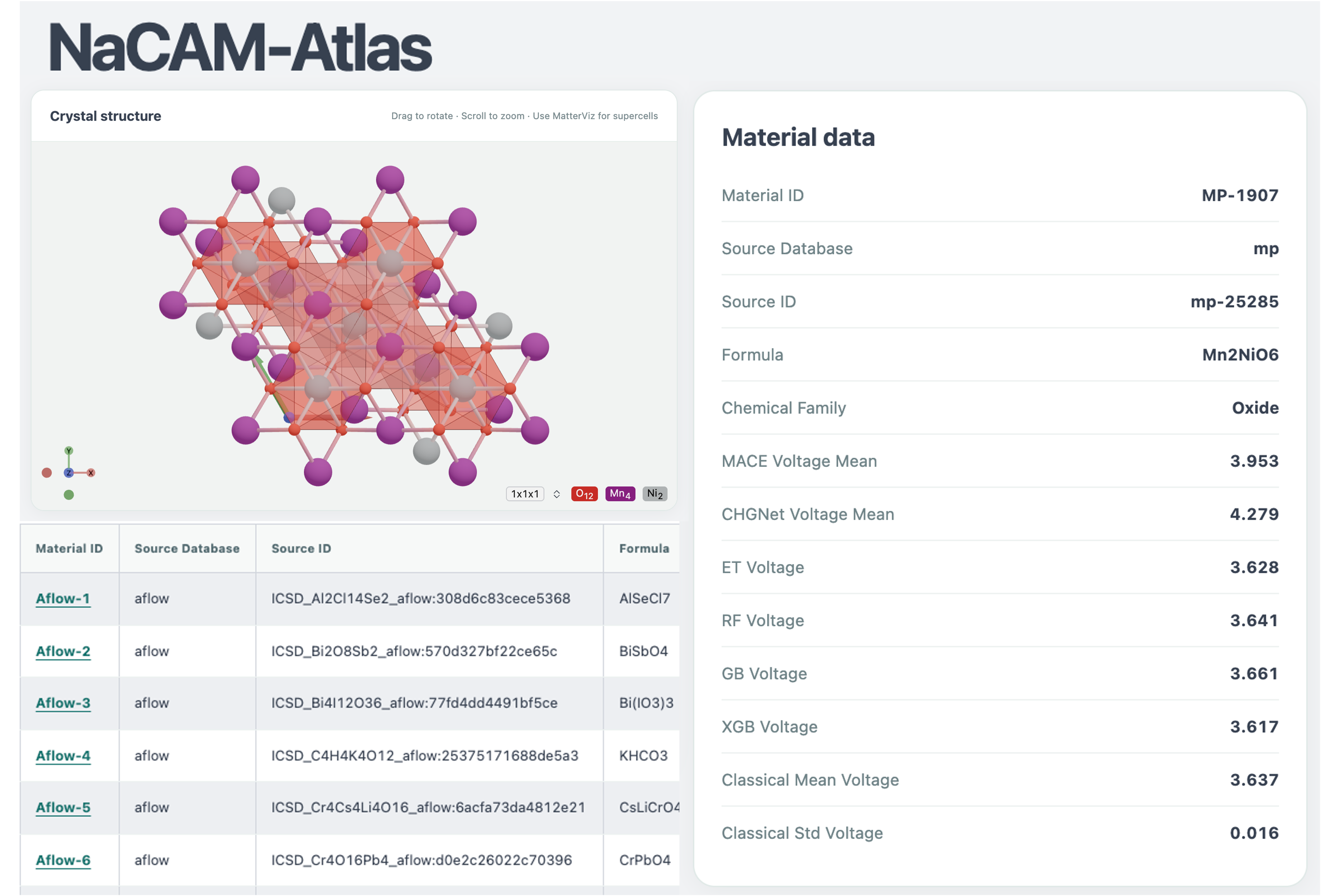}
  \caption{NaCAM-Atlas\cite{NaCAMAtlas2026}, as an interactive database for exploring SIB CAMs.}
  \label{fig:nacam-atlas}
\end{figure*}
Similarly, apparent anion or redox trends have been shown to be substantially affected by uneven chemical sampling across material families.\cite{li2024voltage} Moreover, many TM fluorides undergo conversion rather than topotactic intercalation, unlike phosphate, sulfate, and oxyfluoride hosts.\cite{olbrich2021conversion} Thus, trends among sulfate, phosphate, oxyfluoride, and oxide hosts provide a more robust trend than the heterogeneous fluoride family. Another observed trend in Figure \ref{fig:chemical_family} is that the classical-ensemble model overall predicts higher family-based mean voltages, whereas MACE and CHGNet yield more moderate values. This is consistent with their model designs -- classical tree models work well for interpolation but are less reliable when extrapolating beyond the training domain, whereas MACE and CHGNet use pretrained atomistic representations learned from much larger materials datasets. Therefore, the family-level trends provide further support for the greater extrapolative performance of the pretrained models, which is also consistent with their stronger correlations for the independent 101-structure DFT test set (Figure \ref{fig:correlation}). Moreover, the agreement of MACE and CHGNet with the established cathode chemistry suggests that their pretrained representations capture chemically meaningful trends beyond the original training domain. Overall, the family-resolved trends provide additional chemistry-based validation of the pretrained models and support their use for screening chemically distinct Na-ion cathode candidates beyond the original training domain.

To facilitate exploration of the top-predicted Na-free host materials, we have developed a publicly accessible interactive platform, NaCAM-Atlas\cite{NaCAMAtlas2026} (Figure \ref{fig:nacam-atlas}). Currently, NaCAM-Atlas contains 14,843 structures with Gen-2 MACE-predicted voltages greater than 3 V. All materials in this interactive database are organized by chemical formula, source database, chemical family, predicted voltage, and model uncertainty, based on the disagreement among the classical-ensemble models. Crystal structures can be visualized interactively, and all data, including CIF files, can be downloaded for subsequent analysis. NaCAM-Atlas allows users to distinguish candidates with consistent predictions from those that may require further first-principles evaluations. NaCAM-Atlas also serves as a reusable resource for subsequent model development, followed by experimental synthesis and characterization. The platform can be systematically expanded as additional structures, DFT data, and improved predictive models become available.

\subsection{3.5. First-Principles Calculations of Selected CAMs}
\begin{figure}[h]
  \centering
  \includegraphics[width=0.99\linewidth]{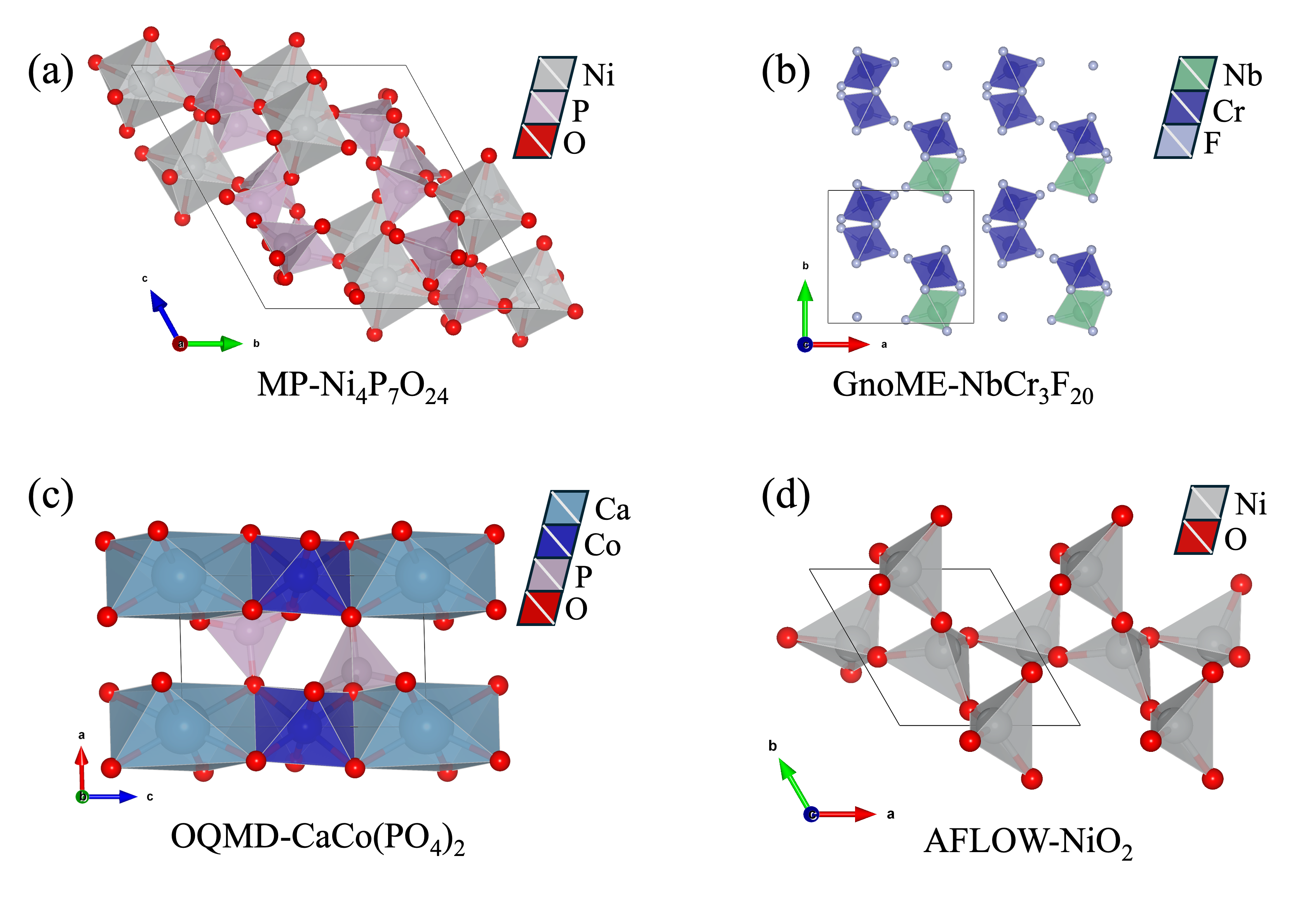}
  \caption{Selected top-ranked CAMs from each database. (a) Ni$_4$P$_7$O$_{24}$ from MP, (b) NbCr$_3$F$_{20}$ from GNoME, (c) CaCo(PO$_4$)$_2$ from OQMD, and (d) NiO$_2$ from AFLOW. See the text for more details.}
  \label{fig:top_4}
\end{figure}

\textbf{3.5. First-Principles Calculations of Selected CAMs}
We finally provide a detailed analysis of selected top-ranked CAMs from each database through explicit, high-throughput DFT calculations of voltage profiles, structural robustness upon sodiation/desodiation, and electronic properties (Figure \ref{fig:top_4}).

\textbf{(a) Ni$_4$P$_7$O$_{24}$.} The charged Ni$_4$P$_7$O$_{24}$ MP structure (ID: mp-26384) consists of distorted NiO$_6$ octahedra interconnected through phosphate framework. Ni-O bond lengths are $\sim$ 1.98--2.17~\AA, while P atoms are tetrahedrally coordinated with P-O distances of $\sim$ 1.49--1.64~\AA. The structure contains both isolated and P-O-P-linked tetrahedral units, creating a strongly connected polyanionic framework. Five Na atoms were inserted with only a 0.40\% change in cell volume, indicating that this highly rigid material can take up Na with minimal expansion. The average DFT-calculated voltage is 4.67~V, while the MACE-predicted voltage is 4.92~V. The charged structure is metallic, and the fully sodiated structure has a small indirect band gap of 0.019~eV. Thus, the material remains electrically conducting throughout the Na insertion. The high average voltage, minimal volume expansion, and preservation of metallic character make Ni$_4$P$_7$O$_{24}$ a highly suitable CAM candidate for SIBs.

\textbf{(b) NbCr$_3$F$_{20}$.} Both Nb and Cr in NbCr$_3$F$_{20}$ from the GNoME database are octahedrally coordinated by F, with Nb-F distances of $\sim$ 1.87--2.11~\AA\ and Cr-F distances of $\sim$ 1.73--1.99~\AA. These metal-fluoride octahedra form edge-sharing units. An NbF$_6$ octahedron is edge-connected to a neighboring CrF$_6$ octahedron, while another edge-sharing unit is formed between two CrF$_6$ octahedra. This framework is less rigid than the previously discussed Ni material from MP. Seven Na atoms were inserted while retaining topotacticity, resulting in an average voltage of 4.63~V, compared to the MACE-predicted voltage of 4.73~V. The fully sodiated structure was found to expand by 18.68\%. This expansion is consistent with the framework's lower connectivity and thus rigidity. The charged structure is semiconducting with a calculated indirect band gap of 1.51~eV. Sodiation reduces the band gap to 0.47 eV, indicating a possible increase in electrical conductivity.

\textbf{(c) CaCo(PO$_4$)$_2$.}
The OQMD CaCo(PO$_4$)$_2$ host consists of distorted CaO$_6$ and CoO$_6$ octahedra interconnected with PO$_4$ tetrahedra. In the desodiated structure, Ca--O and Co--O distances are $\sim$ 2.40--2.44~\AA\ and 1.83--1.93~\AA, respectively, while the P--O distances range from $\sim$ 1.51 to 1.58~\AA. The CaO$_6$ and CoO$_6$ octahedra are edge-sharing, with neighboring Ca and Co polyhedra sharing two O vertices, while the PO$_4$ tetrahedra connect to the metal-oxygen framework through corner-sharing interactions. Formal oxidation-state counting gives Co$^{4+}$ in the desodiated host, which progressively reduces toward Co$^{3+}$ and Co$^{2+}$ upon insertion of one and two Na ions, respectively. The insertion pathway is CaCo(PO$_4$)$_2
\rightarrow$ NaCaCo(PO$_4$)$_2
\rightarrow$ Na$_2$CaCo(PO$_4$)$_2$, with calculated step voltages of 3.50 and 4.26~V, giving an average DFT voltage of 3.88~V. The corresponding average MACE predicted value is 4.24~V. The fully sodiated structure expands by 17.49\%, indicating an appreciable lattice response to Na insertion. Ca and Co remain six-coordinate after sodiation, with Ca--O distances of $\sim$ 2.29--2.42~\AA\ and Co--O distances of $\sim$ 1.99--2.22~\AA, while the inserted Na ions occupy O-coordinated environments with Na--O distances of $\sim$ 2.30--2.70~\AA. The indirect band gap is calculated to be 0.39~eV for the desodiated structure, which increases markedly to 3.23~eV after full sodiation. Thus, although CaCo(PO$_4$)$_2$ retains its polyanionic framework and provides a relatively high average voltage, Na insertion is accompanied by both substantial volume expansion and an increased band gap.

\textbf{(d) NiO$_2$.}
The AFLOW NiO$_2$ host exhibits a markedly different oxide topology, consisting of strongly distorted corner-sharing NiO$_4$ tetrahedra. Ni-based Na$_x$NiO$_2$ cathodes are well established and have been reported to undergo reversible Na extraction/insertion with significant structural evolution during cycling.\cite{vassilaras2013electrochemical} In the desodiated Ni$_3$O$_6$ structure, each Ni is coordinated by four O atoms with Ni--O distances of $\sim$ 1.76--1.83~\AA. Adjacent NiO$_4$ units are connected through individual O vertices rather than shared edges, producing an extended corner-sharing framework. Formal charge counting corresponds to Ni$^{4+}$ in the Na-free composition, consistent with a highly oxidized charged state. The average DFT voltage for this structure is 4.16~V, while MACE predicts 4.73~V. Insertion of two Na ions results in an 18.58\% volume increase. The topotactic framework and fourfold Ni coordination are retained at the final sodiated endpoint, with Ni--O distances of $\sim$ 1.79--1.85~\AA. The inserted Na ions occupy six-coordinated O environments with Na--O distances of $\sim$ 2.25--2.95~\AA. Both the charged and discharged states are zero-band-gap materials. NiO$_2$ therefore retains a metallic character upon Na insertion, but this favorable electronic response is accompanied by substantial lattice expansion.

Our detailed analyses above underscore that high voltage alone may not be sufficient to assess a material's suitability as a CAM. Ni$_4$P$_7$O$_{24}$ combines the highest degree of structural retention with only 0.40\% volume expansion, whereas NbCr$_3$F$_{20}$, CaCo(PO$_4$)$_2$, and NiO$_2$ undergo much larger expansions of 18.68\%, 17.49\%, and 18.58\%, respectively. Their electronic properties also differ substantially: sodiation narrows the band gap of NbCr$_3$F$_{20}$, strongly opens the gap of CaCo(PO$_4$)$_2$, while preserving the metallic behavior in NiO$_2$. These results show the importance of evaluating structural responses and electronic properties alongside predicted voltages when prioritizing ML-screened candidates for further experimental synthesis and characterizations.
\\\\
\section{4. CONCLUSIONS}
\textbf{4. CONCLUSIONS}\newline

In this work, we have developed and evaluated charged-only ML frameworks for at-scale CAM discovery for SIBs. The average voltages are predicted from the desodiated host structures without requiring discharged pairs. We have compared the performance of various classical descriptor-based models, composition- and graph-based NNs, and TL models. Our data show that Random-CV favors the classical models, while LOCO-CV shows lower errors for CHGNet and MACE. This indicates that pretrained foundation models provide an advantage in predicting chemically and structurally unique families. The training set was expanded from 292 to 446 structures using uncertainty-guided DFT labeling. Evaluation on an independent test set showed improved generalization with this expanded test set. 
The Gen-2 models were then used to predict voltages across 840,595 curated structures from the MP, AFLOW, OQMD, and GNoME databases. The contrasting voltage distributions of the MACE model and the classical models showed that different model classes can respond very differently outside their training domain. 
A comprehensive interactive online platform, NaCAM-Atlas, was developed to facilitate exploring SIB CAMs. Finally, selected candidate materials were evaluated through explicit first-principles DFT calculations. It was found that voltage should be considered alongside structural changes and electronic properties to better assess a material's suitability as a CAM for SIBs. Future work will use the developed models to combine voltage predictions with generative and inverse-design approaches to discover novel CAMs beyond known materials.

\begin{suppinfo}
Additional details on the Gen-1 training-set distribution, model performance evaluation metrics, clustering and LOCO validations, Gen-1 and Gen-2 internal validation performances, Gen-2 Pettifor $\oplus$ STR clustering, the high-throughput sodium-insertion workflow used for DFT calculations, and comparison of Gen-1 and Gen-2 CHGNet predictions against the independent DFT test set. The NaCAM-Atlas platform is publicly available at \href{https://huggingface.co/spaces/suchonaakter/NaCAM-Atlas}{https://huggingface.co/spaces/suchonaakter/NaCAM-Atlas}.
\end{suppinfo}

\begin{acknowledgement}
This research was partially supported by the National Science Foundation through Award No. CBET-2302618. Simulations presented in this work used resources from Bridges-2 at Pittsburgh Supercomputing Center through allocations PHY230030P and CHE260021P from the Advanced Cyberinfrastructure Coordination Ecosystem: Services \& Support (ACCESS) program,\cite{access} which is supported by National Science Foundation grants \#2138259, \#2138286, \#2138307, \#2137603, and \#2138296. The authors also acknowledge the HPC center at UMKC for providing computing resources and support.
\end{acknowledgement}

\Addlcwords{is with of in the an a iv v as for on and by to at}

\bibliography{bib}

\begin{tocentry}
\includegraphics[width=0.9\linewidth]{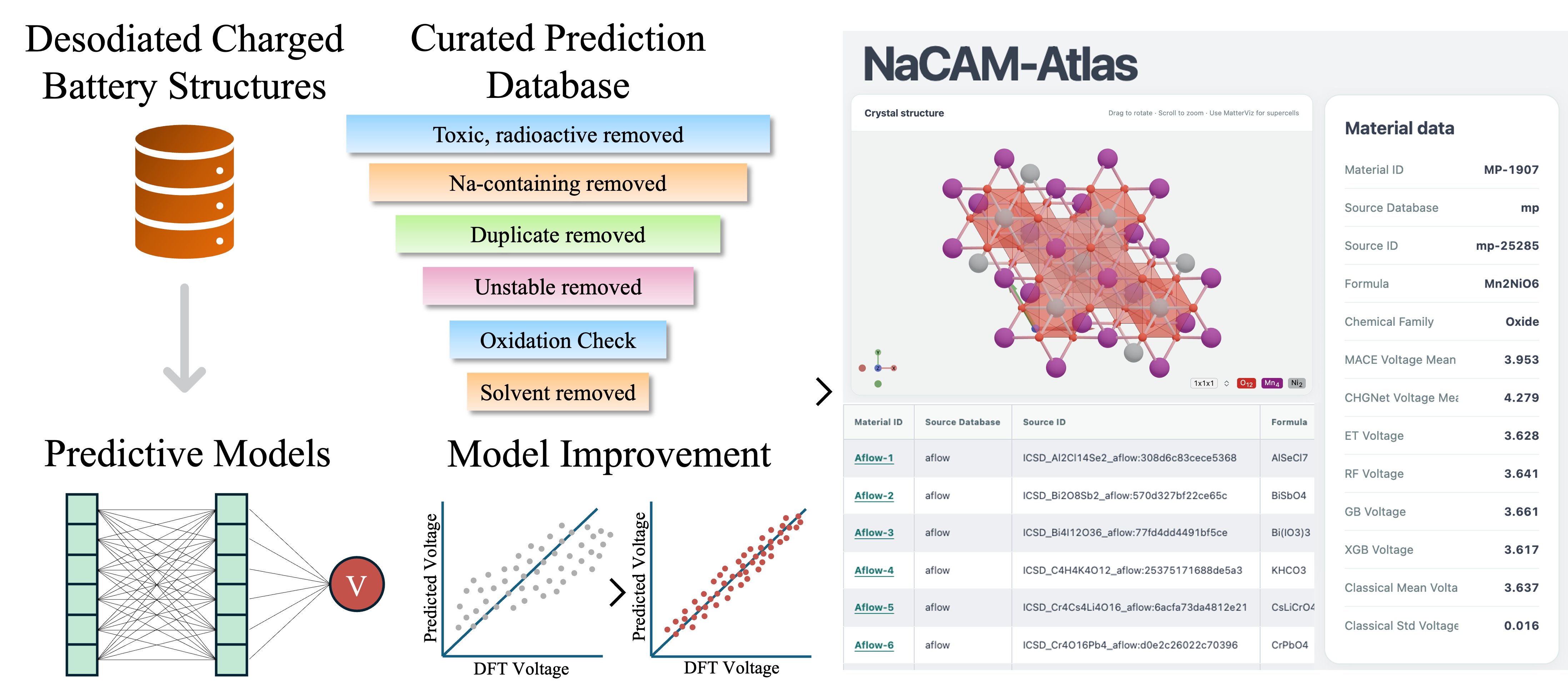}
\end{tocentry}

\end{document}